# Asymmetric Backscattering from the Hybrid Magneto-Electric Meta Particle

Vitali Kozlov[1,a),b)], Dmitry Filonov[1,2,b)], Alexander S. Shalin[2], Ben Z. Steinberg[1] and Pavel Ginzburg[1,2]

[1]*School of Electrical Engineering, Tel Aviv University, Tel Aviv, 69978, Israel*

[2]*ITMO University, St. Petersburg 197101, Russia*

**Abstract**:

The optical theorem relates the total scattering cross-section of a given structure with its forward scattering, but does not impose any restrictions on other directions. Strong backward-forward asymmetry in scattering could be achieved by exploring retarded coupling between particles, exhibiting both electric and magnetic resonances. Here, a hybrid magneto-electric particle (HMEP), consisting of a split ring resonator acting as a magnetic dipole and a wire antenna acting as an electric dipole, is shown to possess asymmetric scattering properties. When illuminated from opposite directions with the same polarization of the electric field, the structure has exactly the same forward scattering, while the backward scattering is drastically different. The scattering cross section is shown to be as low as zero at a narrow frequency range when illuminated from one side, while being maximal at the same frequency range when illuminated from the other side. Theoretical predictions of the phenomena are supported with both numerical and experimental conformations, obtained at the GHz frequency range, and all are in a good agreement with each other. HMEP meta-particles could be used as building blocks for various metamaterials assembling solar cells, invisibility cloaks, holographic masks and more.

---

[a]) vitaliko@mail.tau.ac.il
[b]) V. Kozlov and D. Filonov contributed equally to this work

**Introduction:**

Metamaterials have gained broad interest in the past decade, as they hold the promise for delivering new types of devices[1]. Some of the more notable applications are solar cells [2], invisibility cloaking devices [3], holography [4], opto-mechanics [5],[6] and many more. The basic functionalities of metamaterials are achieved by carefully designing the constitutive elements (meta-atoms) that govern the composite's behavior. Split ring resonators (SRR) and thin wires are often employed as the building blocks in many realizations, owing both to the fact that these structures are well understood theoretically [7], as well as the relative ease of their manufacturing. In order to achieve complex properties, meta-atoms often consist of more than one structure [8],[9],[10],[11]. One of the desirable functionalities that could be achieved with metamaterials is an asymmetric response. For example, asymmetric transmission could find use in a range of applications, such as antireflection coatings [12], light harvesting in solar cells [13], polarization sensitive devises [14],[15] and photonic diodes [16] to name a few.

Scattering characteristics of individual elements could be controlled by engineering their multipolar responses. For example, the so-called Huygen's elements rely on interference between electric and magnetic dipolar responses that suppress the backward scattering [17],[18]. Meta-particles with nonsymmetrical scattering are discussed in details in [19], [20],[21],[22] where a few structures were studied analytically, including the omega, omega-Tellegen and the chiral-moving particle. It was shown that periodic structures constructed from such meta-atoms could be used to create thin films with tunable nonsymmetrical transmission and reflection, e.g. [23] and references therein. Here another example of asymmetric meta-particle is proposed, putting an emphasis on reflection characteristics and its balance with the forward scattering.

The optical theorem [24] relates the forward scattering from an object with its total radar cross-section (RCS) and is a manifestation of the fundamental principle of causality. Remarkably, the theorem favors the forward direction over all the rest and there is no simple relation between the total RCS and the backward scattering, for example. Here a special type of meta-atom, having a symmetric forward and asymmetric backward scattering is studied. The hybrid magneto-electric particle (HMEP), consisting of a SRR and a thin wire (Fig. 1) is considered analytically, numerically and experimentally. The HMEP is shown to have asymmetric backscattering when illuminated by a plane wave from opposite directions.

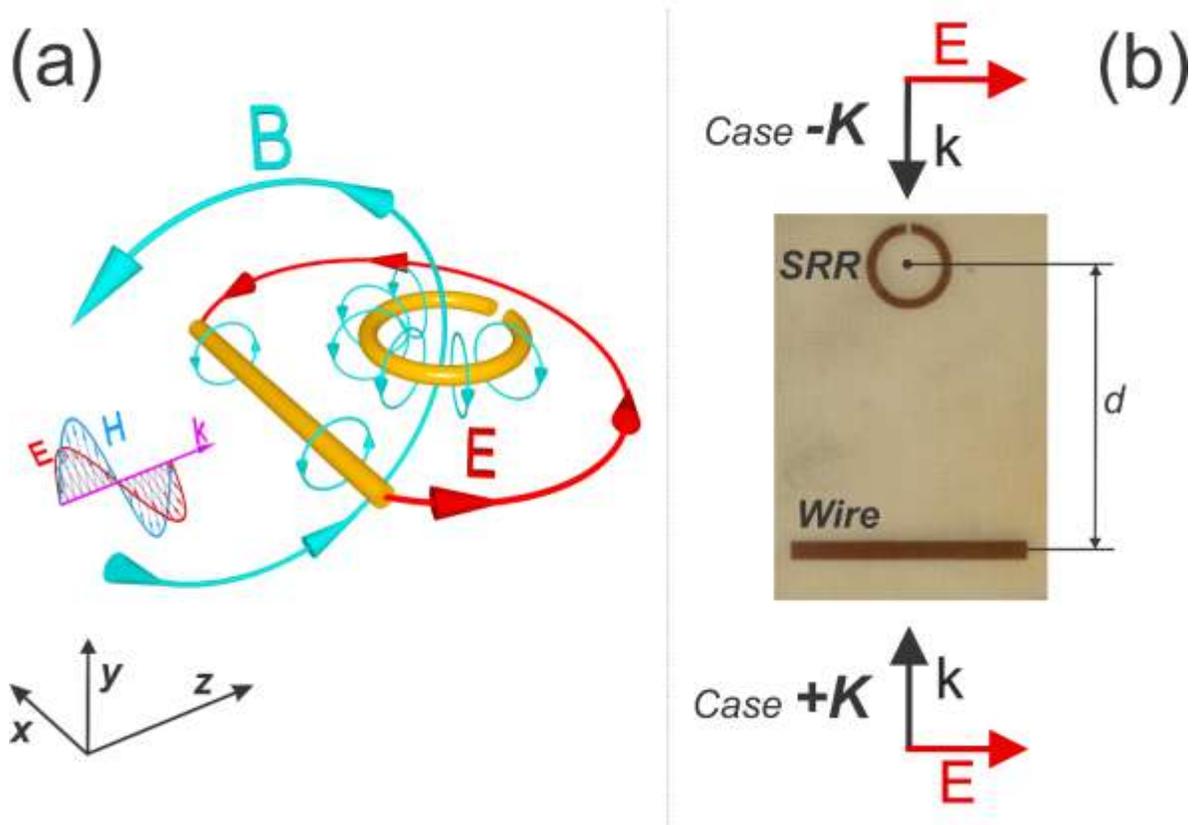

**Figure 1: Hybrid magneto-electric particle (HMEP) geometry (a) Perspective view of the structure and the coupling fields (b) Top view of the fabricated structure (photographic image). In both the forward (case +k) and backward (case -k) propagation the electric field is polarized in the same direction. The split ring resonator (SRR) acts as a magnetic dipole, while the wire as an electric dipole. Coupling between the structures is achieved through the scattered fields. Nonsymmetrical backscattering is achieved owing to the flip in the magnetic field, when the incident wave propagation is reversed.**

It is worth noting, that related structures have been previously analyzed in [25],[26], while theoretical descriptions were reported in [9], where the HMEP was used to construct narrow band filters. Intuitively the asymmetric reflection of the HMEP could be understood by observing Fig.1. Since the polarization of the incident electric field is the same in both the forward and the backward propagation directions, the scattered magnetic field from the wire remains the same when it reaches the SRR. However, the incident magnetic field's polarization switches direction, resulting in constructive or destructive interference with the aforementioned scattered magnetic field in the SRR. This means that the coupling between the two elements is different depending on the direction of propagation.

The letter is organized as follows - first a theoretic framework is derived to show that asymmetric behavior is to be expected for the coupled dipoles. Next, numeric modeling is performed and an optimized structure is compared with experimental results in the GHz range.

**Theory**

Electromagnetic scattering from a structure could be approached with the help of the multipolar decomposition technique [24]. Subwavelength scatterers could often be approximated analytically with a few of the leading (dipolar) terms and the related polarizabilities, which could be calculated directly from the geometry and material composition of the structure. When a structure (meta-atom hereafter) consists of several subwavelength features, the scattering problem could be solved self-consistently by employing the coupled dipoles technique [27],[28].

In the general case, both electric ($p$) and magnetic ($m$) moments could be attributed to each subwavelength scatterer, illuminated by an incident electric $E^{inc}$ and magnetic $H^{inc}$ fields. Matrix equation for a set of discrete coupled dipoles, labeled with sub index '$i$' and situated at points $r_i$, is given in Eq. 1.

$$\begin{bmatrix} p_i \\ m_i \end{bmatrix} = \begin{bmatrix} a^i_{ee} & a^i_{em} \\ a^i_{me} & a^i_{mm} \end{bmatrix} \left( \sum_{\substack{j=1 \\ j \neq i}}^{N} \begin{bmatrix} \epsilon_0^{-1} A(r_i, r_j') & -\mu_0 B(r_i, r_j') \\ B(r_i, r_j') & A(r_i, r_j') \end{bmatrix} \begin{bmatrix} p_j \\ m_j \end{bmatrix} + \begin{bmatrix} E^{inc}(r_i) \\ H^{inc}(r_i) \end{bmatrix} \right), \qquad (1)$$

where $\varepsilon_0$ and $\mu_0$ are the free space permittivity and permeability, $\boldsymbol{a}_{XX'}^i$ is a 3x3 matrix for the electric, magnetic or magneto-electric polarizability of scatterer $i$, while $\boldsymbol{A}(\boldsymbol{r},\boldsymbol{r}')$, $\boldsymbol{B}(\boldsymbol{r},\boldsymbol{r}')$ are the dyadic Green's functions given by:

$$\boldsymbol{A}(\boldsymbol{r},\boldsymbol{r}') = G(R)\left[k^2(\boldsymbol{I}-\boldsymbol{N}) + \left(\frac{1}{R^2}+\frac{jk}{R}\right)(3\boldsymbol{N}-\boldsymbol{I})\right], \\ \boldsymbol{B}(\boldsymbol{r},\boldsymbol{r}') = ck^2 G(R)\left[1+\frac{1}{jkR}\right]\boldsymbol{N}^{(v)}. \tag{2}$$

where $\boldsymbol{I}$ is the 3x3 identity matrix, $R = |\boldsymbol{r}-\boldsymbol{r}'|$, $G(R) = \frac{e^{-jkR}}{4\pi R}$, $\boldsymbol{N} = [\hat{\boldsymbol{n}}^T\hat{\boldsymbol{n}}]$, $\boldsymbol{N}^{(v)} = [(\hat{\boldsymbol{n}}\times\hat{\boldsymbol{x}})^T,(\hat{\boldsymbol{n}}\times\hat{\boldsymbol{y}})^T,\hat{\boldsymbol{n}}\times\hat{\boldsymbol{z}})^T]$, $\boldsymbol{n}=\boldsymbol{r}-\boldsymbol{r}'/|\boldsymbol{r}-\boldsymbol{r}'|$, and k is the free space wave vector. Eq. 1 has a $6N\times 6N$ matrix form ($N$ is the total number of particle scatterers), which could be solved by inversion to find the dipole moments of each scatterer.

In the special case of two simple scatterers and the coordinate system depicted in Fig.1, the SRR's polarizability may be approximated as being dependent only on $\boldsymbol{a}_{mm}^{SRR}$, while the wire's polarizability only on $\boldsymbol{a}_{ee}^{Wire}$:

$$\boldsymbol{a}_{mm}^{SRR} = \begin{pmatrix} 0 & 0 & 0 \\ 0 & a_{SRR} & 0 \\ 0 & 0 & 0 \end{pmatrix}, \boldsymbol{a}_{ee}^{Wire} = \begin{pmatrix} a_{Wire} & 0 & 0 \\ 0 & 0 & 0 \\ 0 & 0 & 0 \end{pmatrix}. \tag{3}$$

The resulting dipolar moments, excited with a plane wave illumination, are our main theoretical result:

$$\boldsymbol{p}^{Wire} = a_{Wire}\frac{1-Sign(k)\mu_0 a_{SRR}\eta^{-1}ve^{-jkd}}{1+\mu_0 a_{Wire}a_{SRR}v^2}\hat{\boldsymbol{x}}, \\ \boldsymbol{m}^{SRR} = a_{SRR}\frac{a_{Wire}v + Sign(k)\eta^{-1}e^{-jkd}}{1+\mu_0 a_{Wire}a_{SRR}v^2}\hat{\boldsymbol{y}}, \tag{4}$$

where $\eta$ is the free space impedance and $v$ is the retarded coupling constant related to the Green's dyadic given by $v = ck^2\frac{e^{-jk|d|}}{4\pi|d|}\left(1+\frac{1}{jk|d|}\right)$. The term $Sign(k)$ explicitly appears in Eq. 4, underlining the dependence of the dipole moments on the propagation direction, where a positive ($+k$) characterizes an incident field propagating in the positive $z$ direction and for a negative $k$ the propagation direction is reversed, as could be seen in Fig.1. The appearance of the $Sign(k)$ term comes from the fact $\boldsymbol{H}^{inc}(\boldsymbol{r}_i)$ in Eq. (1) changes polarization direction when the incident field propagates in the negative $z$ direction with the same polarization of the electric field. This is the core property for the asymmetry effect that could be hardly achieved in a similar fashion with two purely electrical coupled dipoles.

The scattered field is extracted from the dipolar moments and is given by:

$$\begin{bmatrix} \boldsymbol{E}^{sc}(\boldsymbol{r}) \\ \boldsymbol{H}^{sc}(\boldsymbol{r}) \end{bmatrix} = \sum_{j=SRR,Wire} \begin{bmatrix} \epsilon_0^{-1}\boldsymbol{A}(\boldsymbol{r},\boldsymbol{r}_j') & -\mu_0 \boldsymbol{B}(\boldsymbol{r},\boldsymbol{r}_j') \\ \boldsymbol{B}(\boldsymbol{r},\boldsymbol{r}_j') & \boldsymbol{A}(\boldsymbol{r},\boldsymbol{r}_j') \end{bmatrix} \begin{bmatrix} \boldsymbol{p}_j \\ \boldsymbol{m}_j \end{bmatrix}, \quad (5)$$

The RCS is defined as $4\pi L^2 |\frac{E^{sc}(0,0,L)}{E^{inc}(0,0,L)}|^2$ where $L$ is a large distance in comparison with the wavelength and geometrical features of the meta-atom. Direct calculation of the forward and backward scattering could be performed by substituting the relevant numbers in Eq. (5). For example, backward-scattered far-field ($k|z| \to \infty$) is given by:

$$\boldsymbol{E}_B^{SC} = \hat{\boldsymbol{x}} E^{inc} k^2 \frac{\frac{a_{Wire}}{\varepsilon_0} - \eta a_{Wire} a_{SRR} v\left(1 + e^{-jSign(k)kd}\right) - a_{SRR} e^{-jSign(k)kd}}{1 + \mu_0 a_{Wire} a_{SRR} v^2} G(|z|) \quad (6)$$

While the theoretical treatment above is exact for subwavelength structures, additional multipole terms need to be taken into account when the structure's size becomes comparable with the wavelength. Furthermore, the finite dimensions of the particles start playing role in the case of small separation distances ($d$). Instead of using the multipolar expansion, which may become involved for numerous terms, the scattered field could be calculated numerically. This approach is undertaken ahead using CST Microwave Studio. In order to apply Eq. 6 and calculate the asymmetry factor, wire and SRR polarizabilities ($a_{Wire}, a_{SRR}$) should be obtained first. For this purpose full wave simulations on individual elements (wire and SRR) were performed and numerical values of polarizabilities were extracted from backscattered fields (e.g. $a_{Wire}$ was obtained from Eq. 6 by putting , $a_{SRR} = 0, d = 0$). Fig. 2. summarizes the results for the structure in a free space (dipole – thin strip 22.3x2mm, SSR – outer diameter – 8.3mm, inner 6.3mm, the gap – 0.5mm). Since substrates could introduces secondary contributions, such as resonance shifts, image secondary sources and losses, they were omitted from those theoretical investigations, aiming on underlining the main effect. The asymmetry factor was defined as a ratio between the difference and the sum of backward scatterings at opposite direction of incidence ($\pm k$). This factor is bounded between $\pm 1$ similar to visibility coefficients, characterizing interference fringes.

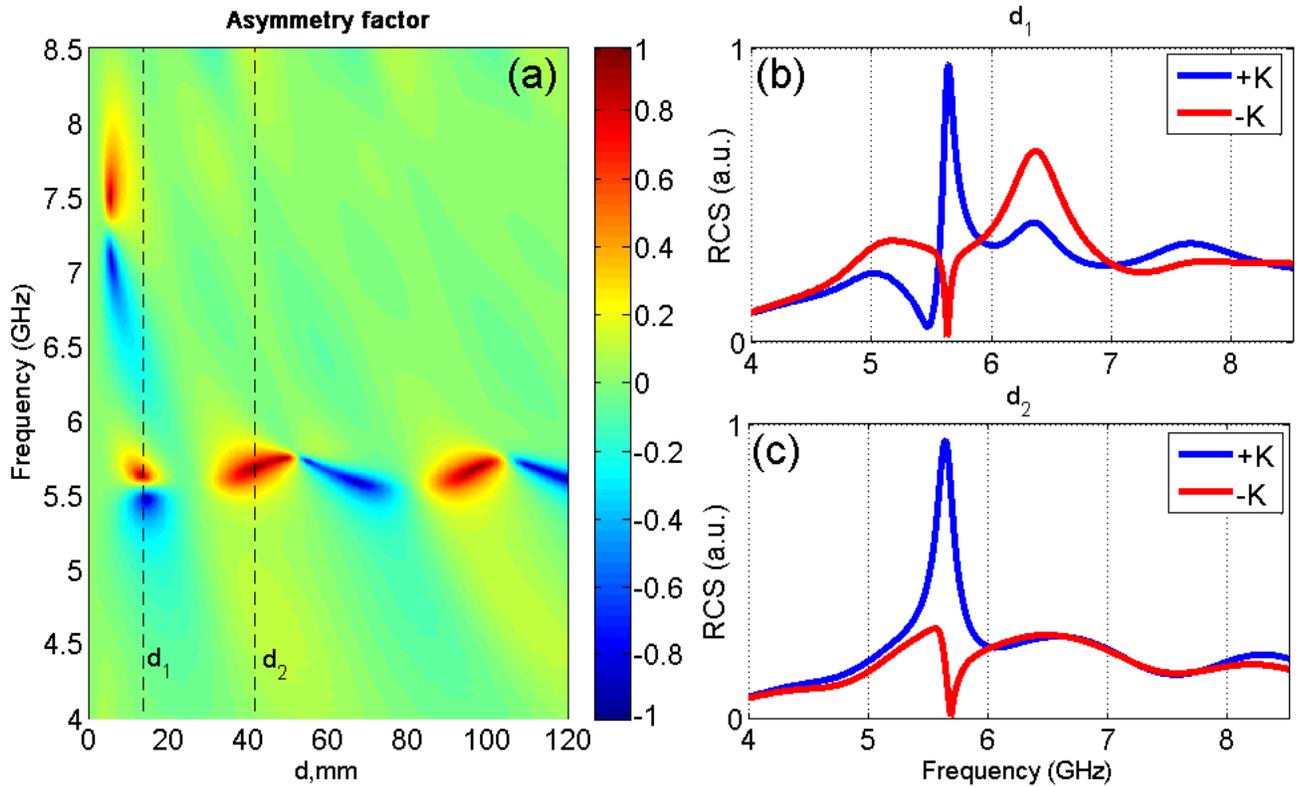

**Figure 2: Asymmetry in backward scattering from HEMP particles.** (a) Asymmetry factor (color map) as the function of frequency of the incident wave and the separation distance between the dipole and the SRR (both assumed

to be point particles in comparison to the separation distance), forming the HEMP. (b), (c) Radar cross-section spectra of HEMPs with fixed dimensions ($d_1 = 13.6 \, mm$, $d_2 = 41.6 \, mm$). Slices are shown on panel (a). Blue and red curves correspond to opposite direction of incidence, indicated on Fig. 1. Sizes of the HEMP elements - (dipole – thin strip of 22.3x2mm, SSR – outer diameter – 8.3mm, inner 6.3mm, the gap – 0.5mm).

The asymmetry factor (Fig. 2(a)) shows a strong dependence on the system's parameters – illumination frequency and the separation distance between the ring and the wire. First, the strong asymmetry could be achieved with subwavelength dimensions (Fig. a(b)). This property is unobtainable without employing a combination of electric and magnetic resonances – that is the unique property of the HEMP particle. The asymmetry also preserves at wavelength-comparable separation distances – Fig. 2(c)). An additional very remarkable feature is the strong sensitivity of the asymmetry factor on the separation distance – as it can be seen from Fig. 2(a)) a small change in $d$ (on the scale of millimeters – less than one tenth of the wavelength) could flip the sign of the asymmetry coefficient. Such a high sensitivity could be employed for e.g. sensing applications.

**Numerical Analysis and Microwave Experiment**

The photographic image of the fabricated HMEP particle appears in Fig. 1(b). The design of the constitutive elements is based on the standard copper strips, printed on a dielectric substrate (FR4 fiberglass, having $\varepsilon \approx 4.4$). In order to investigate asymmetric backscattering from HMEP, its basic components (the SRR and the wire) are analyzed first. Numerical simulations were performed with the frequency domain finite element method, implemented in CST Microwave Studio, aiming to emulate the experimental layout as close as possible, including the substrate effects. In order to implement the plane wave excitation in the laboratory conditions and measure the signal scattered to both the forward and backward directions, a pair of rectangular linearly polarized wideband horn antennas (1–18 GHz) were employed. Antennas were connected to coaxial ports of a vector network analyzer (Agilent E8362B), able to extract both the amplitude and phase of the received signals. The experiment was conducted in an anechoic chamber where the HMEP was positioned in the far-field of the antennas. The distance from the HMEP to the transmitting and receiving antennas was approximately 1.5 m. The backscattering was extracted from the measured complex-valued signals, obtained in several steps in order to eliminate instrumental responses. The calibration was performed with a large area metallic mirror, having a reflection coefficient equal to -1. All the reflected signals were normalized accordingly to factorize the impact of the measurement apparatus.

Fig. 3 shows the RCSs of the separate elements appearing in Fig. 1. It can be seen that the two are resonant at different frequencies with the SRR having a narrower resonance, which is in good agreement with experimental data. It should be noted, that individual elements were taken from the geometry of the combined particle, which was optimize to deliver the highest asymmetric response. As the result, individual elements do not resonate at the same frequency, as could be seen from Fig. 3. However, the mutual coupling between constitutive elements will cause the complex particle to exhibit one strong resonance. Additional wavy behavior, which appears on the experimental data (Fig. 3(b)), is attributed to additional Fabry-Perot resonances between the sample and the horn antennas and is quite common in experiments of this kind.

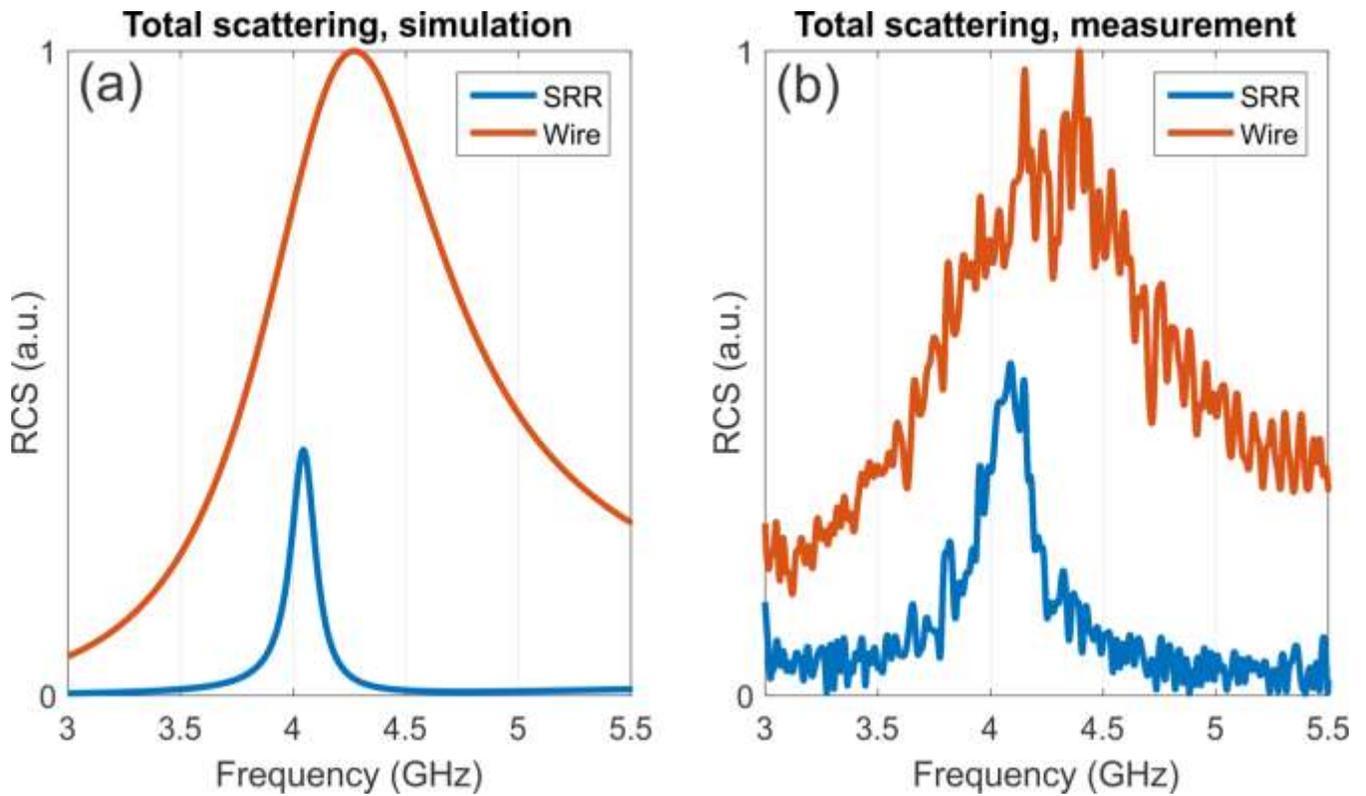

**Figure 3: Scattering cross sections of the separate constitutive elements - split ring resonator (blue curves) and wire (red curves). (a) Simulated (b) Experimental total scattering cross section, as a function of frequency.**

Fig.4 shows the simulated forward and total scattering cross sections of the entire structure, depicted in Fig.1. Solid blue and dashed red lines correspond to opposite directions of the incident plane wave. It can be seen that for both positive and negative *k* the scattering measures remains the same. Those numerical results underline the validity of the optical theorem, relating forward scattering to the RCS. As it could be explicitly seen, the theorem is satisfied and no distinction on the propagation direction could be made.

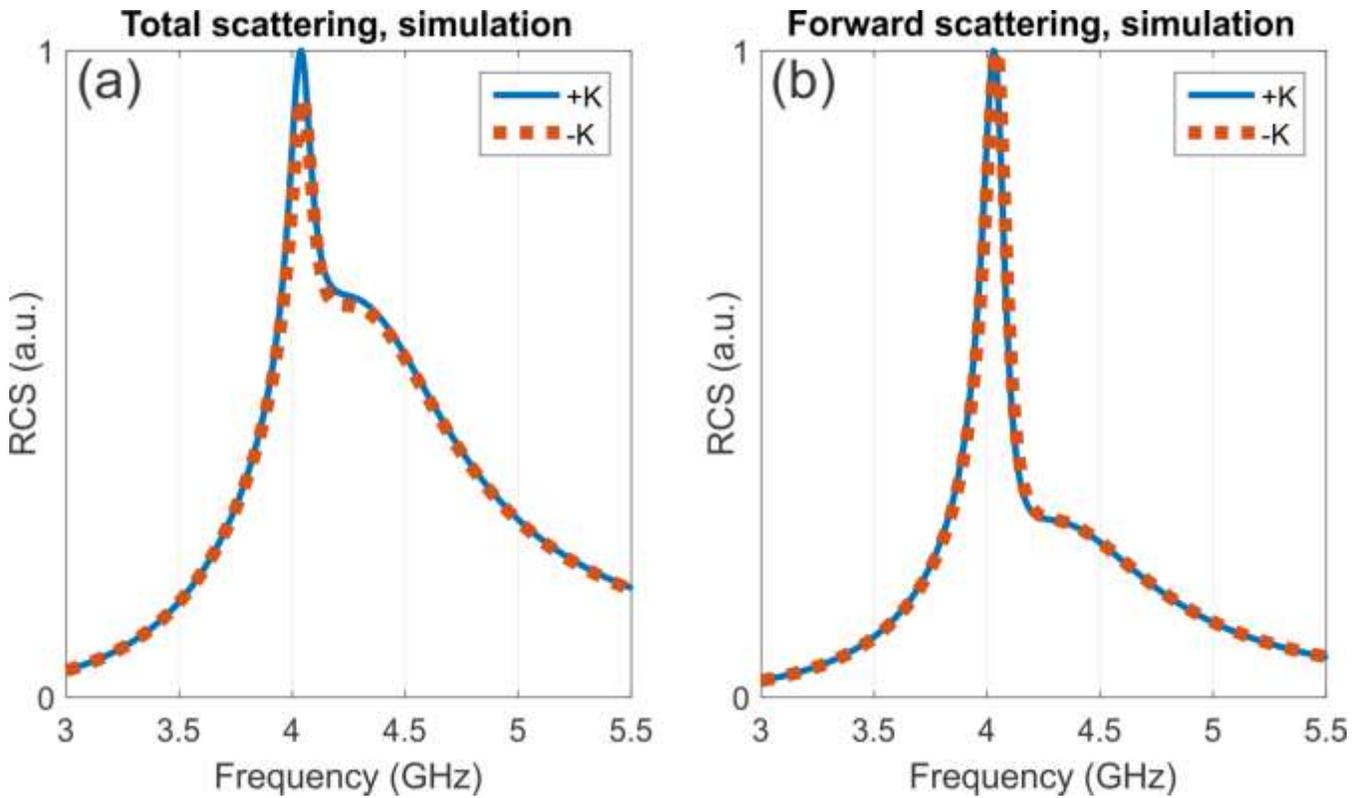

**Figure 4: (a) Simulation of the total scattering cross section for forward and backward incident field propagations. (b) Simulation of the forward scattering. For both propagation directions the scattering appears the same.**

Fig. 5 is the main result, showing that the backscattering on the z axis at a large distance away from the HMEP, strongly depends on the direction of incident field propagation. In the forward direction (+k) the field is strongly reflected around 4GHz while in the backward direction (-k) there is almost no backscattering at all in the same frequency range. In order to satisfy optical theorem, the scattered radiation should be redistributed along other directions in space. This behavior is explicitly shown in the inset to Fig.4(a), where polar plots of scattering diagrams appear. As it may be seen, in the +k case the backward scattering is quite significant, it vanishes once the incidence direction is flipped. Scattering to sides of the HEMP increases on the expense of the backscattering. Fig. 5(b) shows the experimental data on the asymmetric scattering and it is in an excellent agreement with the numerical investigations. Furthermore, the numerical results (geometry that includes the substrate) and theoretical predictions (Fig. 2) are in a very good agreement either – spectral behavior of curves is alike apart from a resonance shift, caused by the substrate. This effect suggests this meta-particle to be a viable building block for different metamaterials that could benefit from asymmetric response to different incident field propagation directions.

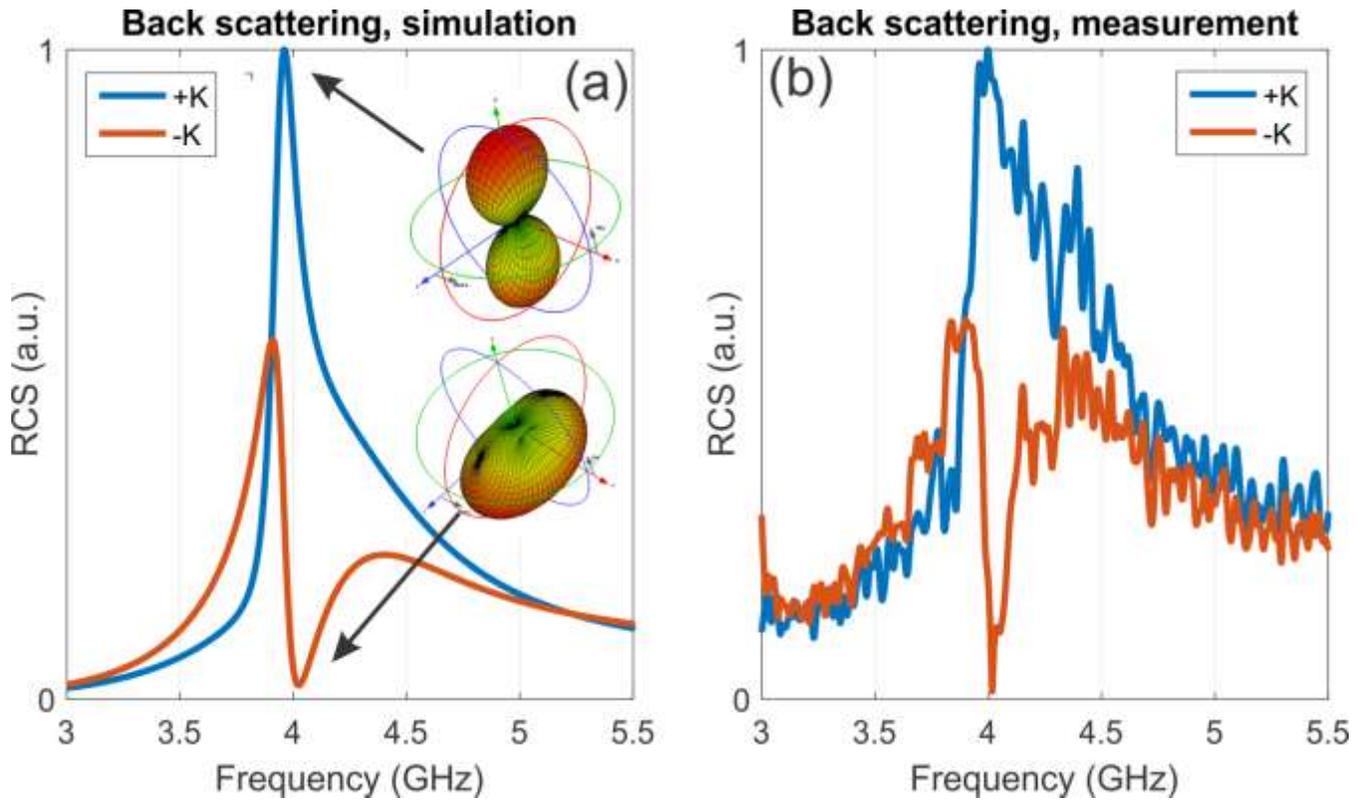

**Figure 5: Backscattered field from the HMEP for the different directions of the incident plane wave. (a) Simulated backscattering with the insert radiation patterns at 4GHz. Insets – scattering patterns (polar plots) for different directions of incidence. (b) Experimental measurement of the backscattered field for both propagation directions of the incident field.**

## Conclusions

The HMEP meta-particle was analyzed analytically, numerically, and experimentally and was shown to have strong asymmetric response – being illuminated from opposite directions, it has completely symmetric forward scattering, while maximum asymmetry was achieved in the backward direction. The backscattering from the particle may be as low as 0 for one direction of the incident wave, while being maximal at the same frequency range for the opposite direction of propagation. It should be underlined that the optical theorem is not violated, and the scattered signal is merely redistributed between the other directions in space.

One of the fundamental properties of Maxwell's equations is that they are frequency scalable – this means that scattering from meta-atoms designed for the optical frequencies would be analogous to scattering from a similar meta-atom designed for radio frequencies (RF), provided that the material parameters $\mu$ and $\varepsilon$ are the same [29]. This scalability enables emulation of optical metamaterials, which are often difficult to fabricate due to their small dimensions, with the relatively easily manufacturable metamaterials in the RF range [30],[31]. Therefore, the proposed structure could be employed for approaching a range of applications in the optical domain. Consequently, HMEP could be used as a building block for asymmetric metamaterials assembling parts of solar cells, optical diodes and more.

**Acknowledgements:**

This work was supported, in part, by TAU Rector Grant and German-Israeli Foundation (GIF, grant number 2399) and by the Russian Fund for Basic Research within the project 16-52-00112. The analytical calculation of multipolar terms of meta-particle has been supported by the Russian Science Foundation Grant No. 16-12-10287. A.S.S. acknowledges the support of the President of Russian Federation in the frame of Scholarship SP-4248.2016.1 and the support of Ministry of Education and Science of the Russian Federation (GOSZADANIE 2014/190). The authors acknowledge Dr. Tal Ellenbogen (Tel Aviv University) for discussions.

## References:

[1]     R. W. Z. Nader Engheta, *Metamaterials: Physics and Engineering Explorations*. Wiley-IEEE Press.

[2]     Y. Wang, T. Sun, T. Paudel, Y. Zhang, Z. Ren, and K. Kempa, "Metamaterial-Plasmonic Absorber Structure for High Efficiency Amorphous Silicon Solar Cells," *Nano Lett.*, no. 12, pp. 440–445, 2012.

[3]     D. Schurig, J. J. Mock, B. J. Justice, S. A. Cummer, J. B. Pendry, A. F. Starr, and D. R. Smith, "Metamaterial Electromagnetic Cloak at Microwave Frequencies," *Science (80-. ).*, vol. 314, no. 5801, pp. 977–80, 2006.

[4] S. Larouche, Y. Tsai, T. Tyler, N. M. Jokerst, and D. R. Smith, "Infrared metamaterial phase holograms," *Nat. Mater.*, vol. 11, no. 5, pp. 450–454, 2012.

[5] A. S. Shalin, S. V. Sukhov, A. A. Bogdanov, P. A. Belov, and P. Ginzburg, "Optical pulling forces in hyperbolic metamaterials," *Phys. Rev. A*, vol. 91, no. 6, p. 063830, Jun. 2015.

[6] A. A. Bogdanov, A. S. Shalin, and P. Ginzburg, "Optical forces in nanorod metamaterial.," *Sci. Rep.*, vol. 5, p. 15846, Jan. 2015.

[7] B. Sauviac, C. R. Simovski, and S. A. Tretyakov, "Double Split-Ring Resonators : Analytical Modeling and Numerical Simulations," *Electromagnetics*, vol. 24, no. 5, pp. 317–338, 2016.

[8] O. Sydoruk, "Tailoring the near-field guiding properties of magnetic metamaterials with two resonant elements per unit cell," *Phys. Rev. B*, vol. 73, no. 224406, pp. 1–12, 2006.

[9] Y. Hadad and B. Z. Steinberg, "Electrodynamic Synergy of Micro-properties and Macro-structure in Particle Arrays," *URSI Int. Symp. Electromagn. Theory*, pp. 680–683, 2010.

[10] A. O. Karilainen and S. A. Tretyakov, "Isotropic Chiral Objects With Zero Backscattering," *IEEE Trans. Antennas Propag.*, vol. 60, no. 9, pp. 4449–4452, 2012.

[11] D. S. Filonov, A. P. Slobozhanyuk, P. A. Belov, and Y. S. Kivshar, "Double-shell metamaterial coatings for plasmonic cloaking," *Phys. Status Solidi - Rapid Res. Lett.*, vol. 6, no. 1, pp. 46–48, 2012.

[12] K. X. Wang, Z. Yu, S. Sandhu, V. Liu, and S. Fan, "Condition for perfect antireflection by optical resonance at material interface," *Optica*, vol. 1, no. 6, p. 388, Dec. 2014.

[13] K. Tvingstedt, S. D. Zilio, O. Inganäs, and M. Tormen, "Trapping light with micro lenses in thin film organic photovoltaic cells," *Opt. Express*, vol. 16, no. 26, 2008.

[14] Y. Liu and X. Zhang, "A new frontier of science and technology," *Chem. Soc. Rev*, vol. 40, pp. 2494–2507, 2011.

[15] A. P. Slobozhanyuk, P. V. Kapitanova, D. S. Filonov, D. A. Powell, I. V. Shadrivov, M. Lapine, P. A. Belov, R. C. McPhedran, and Y. S. Kivshar, "Nonlinear interaction of meta-atoms through optical coupling," *Appl. Phys. Lett.*, vol. 104, no. 1, pp. 1–5, 2014.

[16] I. V Shadrinov, V. A. Fedotov, D. A. Powell, Y. S. Kivshar, and N. I. Zheludev, "Electromagnetic wave analogue of an electronic diode," *New J. Phys.*, vol. 13, no. 033025, 2011.

[17] A. E. Krasnok, A. E. Miroshnichenko, P. A. Belov, and Y. S. Kivshar, "All-dielectric nanoantennas," *Proc. SPIE*, vol. 8806, 2013.

[18] A. O. Karilainen, P. Alitalo, and S. A. Tretyakov, "Chiral Antenna Element as a Low Backscattering Sensor," pp. 1865–1868, 2011.

[19] Y. Ra, S. Member, V. S. Asadchy, and S. A. Tretyakov, "Total absorption of electromagnetic waves in ultimately thin layers," *IEEE Trans. Antennas Propag.*, vol. 61, no. 9, pp. 4606–4614, 2013.

[20] Y. Ra, S. Member, V. S. Asadchy, and S. A. Tretyakov, "Tailoring Reflections From Thin Composite Metamirrors," *IEEE Trans. Antennas Propag.*, vol. 62, no. 7, pp. 3749–3760, 2014.

[21] Y. Ra, V. S. Asadchy, and S. A. Tretyakov, "One-way transparent sheets," *Phys. Rev. B*, vol. 89, no. 075109, pp. 1–7, 2014.

[22] V. V. Klimov, I. V. Treshin, A. S. Shalin, P. N. Melentiev, A. A. Kuzin, A. E. Afanasiev, and V. I. Balykin, "Optical Tamm state and giant asymmetry of light transmission through an array of nanoholes," *Phys. Rev. A*, vol. 92, no. 6, p. 063842, Dec. 2015.

[23] C. Pfeiffer and A. Grbic, "Metamaterial Huygens' Surfaces: Tailoring Wave Fronts with Reflectionless Sheets," *Phys. Rev. Lett.*, vol. 110, no. 19, p. 197401, May 2013.

[24] J. D. Jackson, "Classical Electrodynamics, 3d ed." p. 500.

[25] S. Gupta, L. J. Jiang, S. Member, and C. Caloz, "Magneto-Electric Dipole Antenna Arrays," *IEEE Trans. Antennas Propag.*, 2014.

[26] C. R. Simovski, P. A. Belov, S. Member, S. He, and S. Member, "Backward Wave Region and Negative Material Parameters of a Structure Formed by Lattices of Wires and Split-Ring Resonators," *IEEE Trans. Antennas Propag.*, vol. 51, no. 10, pp. 2582–2591, 2003.

[27] L. Novotny and H. Bert, *Principles of Nano-Optics*. 2006.

[28] D. Markovich, K. Baryshnikova, A. Shalin, and A. Samusev, "Enhancement of artificial magnetism via resonant bianisotropy," *Sci. Rep.*, vol. 6, no. 22546, pp. 1–8, 2016.

[29] B. Hopkins, D. S. Filonov, S. B. Glybovski, and A. E. Miroshnichenko, "Hybridization and the origin of Fano resonances in symmetric nanoparticle trimers," *Phys. Rev. B - Condens. Matter Mater. Phys.*, vol. 92, no. 4, pp. 1–10, 2015.

[30] D. S. Filonov, A. S. Shalin, P. A. Belov, and P. Ginzburg, "Emulation of complex optical phenomena with radio waves: Tailoring scattering characteristics with wire metamaterial," in *2015 IEEE International Conference on Microwaves, Communications, Antennas and Electronic Systems (COMCAS)*, 2015, pp. 1–2.

[31] P. V Kapitanova, P. Ginzburg, F. J. Rodríguez-Fortuño, D. S. Filonov, P. M. Voroshilov, P. A. Belov, A. N. Poddubny, Y. S. Kivshar, G. A. Wurtz, and A. V Zayats, "Photonic spin Hall effect in hyperbolic metamaterials for polarization-controlled routing of subwavelength modes.," *Nat. Commun.*, vol. 5, p. 3226, Jan. 2014.